# Spin-chiral electron-phonon coupling in metallic strontium titanate

N. Somun[1], L. Rogić[1], I. Khayr[2], E. Tafra[1], M. Basletić[1], A. Kundu[3], J. Dzian[4], F. Le Mardelé[4], M. Orlita[4], M. Greven[2], B. Büchner[5,6], A. Alfonsov[5], M. N. Gastiasoro[7*], A. Klein[8*], D. Pelc[1,2*]

[1]Department of Physics, Faculty of Science, University of Zagreb, HR-10000 Zagreb, Croatia

[2]School of Physics and Astronomy, University of Minnesota, MN-55455 Minneapolis, USA

[3]Asia Pacific Center for Theoretical Physics (APCTP), Pohang, Gyeongbuk 37673, Korea

[4]Laboratoire National des Champs Magnétiques Intenses, 38000 Grenoble, France

[5]Leibniz Institute for Solid State and Materials Research Dresden, Helmholtzstr. 20, 01069 Dresden, Germany

[6]Institute for Solid State and Materials Physics and Würzburg-Dresden Cluster of Excellence, TU Dresden, 01062 Dresden, Germany

[7]Donostia International Physics Centre, 20018 Donostia-San Sebastian, Spain

[8]Department of Physics, Bar-Ilan University, Ramat Gan, Israel

*correspondence to: maria.ngastiasoro@dipc.org, avraham.klein@biu.ac.il, dpelc@phy.hr

**Electron-phonon coupling (EPC) – the interaction between conduction electrons and quantized atomic vibrations – plays a central role in condensed matter physics and determines some of the most important properties of materials, such as electrical resistivity and superconductivity. Conventionally, EPC is assumed to be induced by the ionic electrostatic background, and electronic spin plays no role in the process. In stark contrast with this view, here we uncover a direct spin-mediated coupling mechanism between electrons and transverse polar phonons in a metal. Using far-infrared light absorption measurements of the model system $SrTiO_3$ in a magnetic field, we observe a strong spin-mediated EPC that is quantitatively consistent with recent theoretical predictions [1-8], and that generates chiral phonon modes with large effective magnetic moments. The extracted coupling strength is in good agreement with *ab initio* estimates [6] and sufficiently high to explain superconductivity in $SrTiO_3$, thereby resolving a long-standing conundrum [9,10]. Spin-chiral EPC should generically appear in all metals with polar phonons, and the present work could be of relevance to spintronics applications [11] and to uncovering the origins of superconductivity in layered materials [12,13], metals with Dirac points in their electronic dispersions [14,15], and nearly ferroelectric superconductors [16-21].**

EPC is one of the fundamental interactions in materials: it is not only the microscopic origin of resistivity and superconductivity in most metals, but also gives rise to complex phases such as charge-density waves. The coupling is strongest if the electrons can interact with single phonons, while multi-phonon interactions are typically much weaker. The standard one-phonon process results from the gradient of the effective potential felt by the electrons and is active only for modes with longitudinal polarization. Recently, a novel one-phonon coupling mechanism between electrons and *transverse* polar phonons was proposed, which involves electronic spin-orbit coupling and is a dynamic analogue of the well-known static Rashba effect in noncentrosymmetric metals [1-8]. This Rashba-like EPC has been suggested to be most prominent in systems with soft polar phonon modes, but it should appear in any metal with phonons that dynamically break inversion symmetry. Moreover, the spin-orbit-mediated EPC has been predicted to lead to a hybridization between phonons and electronic spin-flip excitations [4,22], which imparts a magnetic moment and well-defined handedness to the phonon-like modes in external magnetic fields. Such chiral phonons are of great interest, both for their unusual fundamental properties [23-27] and for possible advanced applications, e.g., the manipulation of emergent magnetism in non-magnetic materials and the generation of spin-polarized currents [11].

Strontium titanate ($SrTiO_3$, STO) is a remarkable quantum material and an ideal model system in which to experimentally detect the proposed Rashba-like coupling. Pristine STO is a band insulator and extremely close to a ferroelectric instability, but true long-range ferroelectric order does not appear at any temperature. Upon doping with tiny concentrations of mobile electrons, STO undergoes an insulator-metal transition [28], and the low-density metallic state has been the subject of long-standing interest due to its unusual properties, including unconventional electronic transport [29-31] and superconductivity [32]. The ratio of the superconducting transition temperature, $T_c$, to the Fermi temperature is among the highest of all known superconductors, and the microscopic mechanism responsible for superconducting pairing remains debated after decades of intense study [9,10]. While a number of experimental [33-37] and theoretical [38,39] investigations have suggested that the soft transverse optic (TO) phonon mode associated with ferroelectric fluctuations plays a pivotal role, the nature of the electron-phonon coupling has not been resolved. Since the conventional gradient coupling is zero, two theoretical scenarios have emerged: a second-order, two-phonon coupling mechanism [40-43], and the aforementioned Rashba-like interaction [1-6]. The experimental determination of the respective coupling strengths is therefore crucial to discern the superconducting pairing mechanism in metallic STO and related materials where soft transverse phonons are present, such as STO/LAO interfaces [44] and $KTaO_3$ heterostructures [16-18,21].

Here we utilize far-infrared light to probe the influence of a magnetic field on soft polar phonons in STO by leveraging the recent development of a sensitive absorption spectrometer for sub-terahertz frequencies [45]. We detect strong spin-phonon hybridization in metallic samples, in quantitative agreement with theoretical predictions for the Rashba-like EPC, and we use circular dichroism to demonstrate that the spin-hybridized phonon-like modes acquire a significant magnetic moment. In contrast, the magneto-optical response of pristine, insulating STO is found to be negligible, which implies that the dramatic magnetic field-induced changes originate entirely from the interaction between phonons and conducting electrons.

Optical spectroscopy can only probe phonons with very small momenta, i.e., close to the centre of the first Brillouin zone. At low temperatures, a slight tetragonal distortion of the STO cubic perovskite structure results in two non-degenerate soft polar TO modes [46]. The $A_{2u}$ phonon is polarized along the crystallographic $c$-axis and has higher energy, while the polarization of the softer $E_u$ mode lies in the basal plane of the tetragonal structure (Fig. 1a). We always observe a combination of both modes because of the existence of three distinct tetragonal domains. In order to estimate the phonon magnetic moment in the absence of conduction electrons, we first consider the magneto-optical response at frequencies close to the $E_u$ mode in insulating STO, in magnetic fields up to 30 T (Fig. 1b). Within resolution, we do not observe any field-induced changes in the optical absorption, which places an upper limit of ~$0.01\mu_B$ on the effective phonon magnetic moment, where $\mu_B$ is the Bohr magneton (see Methods for details). In contrast, we find a significant magneto-optic response in metallic samples. Figure 1c shows the optical absorption of an oxygen-vacancy doped (OVD) STO crystal with charge carrier density $n \approx 2{\cdot}10^{18}$ cm$^{-3}$ (labelled $OVD_a$). At this low carrier concentration, electronic features such as plasma absorption edges (arrow in Fig. 1c) lie well below the energy of $E_u$ phonon, which results in an exceptionally clear hybridized phonon-spin-flip excitation. Assuming that the atomic displacement pattern of the phonon mode is dominated by a simple antiphase oscillation of the oxygen cage and the titanium atom (Fig. 1a), a fit of the theoretical prediction to the lower-frequency branch yields a Rashba-like EPC strength $\tau_R$ = 400±80 meV/Å (see Methods for details), within a factor of two of frozen-phonon *ab initio* estimates [5,6]. The energy of the $E_u$ phonon at the experimental temperature of 0.7 K is roughly 0.6 meV, significantly below 0.9 meV obtained for undoped STO in hyper-Raman measurements at about 10 K [46] (see also Methods). This mode therefore softens even more upon cooling than previously known, which further boosts the electron-phonon interaction [6].

Intriguingly, the data show evidence for additional structure in the phonon spectrum above the main branch, i.e., overtones, close to the phonon-spin-flip anticrossing point: at least two minima are seen in the raw spectra (Fig. 1d), in increments of about 0.1 meV above the $E_u$ phonon. The only excitations in the system at this energy scale are the screened plasma oscillations, and the overtone features could therefore be interpreted as the sum of the phonon and plasmon energies. While the lowest-order theoretical calculations do not predict such features, they could arise in a more complete treatment of the hybridized phonon-plasmon excitations in the presence of the Rashba-like coupling and phonon anharmonicity.

The theory of the Rashba-like coupling makes two specific predictions for the dependence of the spin-phonon hybridization on light polarization, which we test explicitly. First, the effect is expected to be absent if the light is linearly polarized along the external magnetic field, which we confirm in linear polarization experiments on a sample with carrier concentration $7.2{\cdot}10^{18}$ cm$^{-3}$ (sample $OVD_b$). Phonons polarized along the magnetic field are not affected, while phonons with perpendicular polarization show spin-phonon hybridization (Fig. 2a,b). Second, the interaction leads to the appearance of well-defined chirality of the hybridized phonon-spin-flip excitations, that can be tuned with magnetic field. In order to directly detect the phonon chirality, we measure the difference of absorption of left and right circularly polarized light, known as magnetic circular dichroism (CD), close to the zero-field $E_u$ phonon energy. The CD response is considerable, and shows the expected symmetry with applied field: the field dependence of the relative absorption is inverted for left- and right-circular polarizations (Fig. 2c), i.e., a simultaneous change of the sign of the magnetic field and the handedness of the light leaves the signal unchanged. The measured CD also agrees well with the theoretical prediction (Fig. 2d; see Methods for details).

In order to determine the dependence of the electron-phonon coupling strength on the carrier concentration, we perform energy- and field-dependent magneto-optical measurements on three additional samples: $OVD_b$ (Fig. 3a, same sample as in Fig. 2) and two samples doped via 0.1% (Fig. 3b) and 0.2% (Fig. 3c) niobium substitution, respectively. The spectra show two distinct, overlapping types of features: plasma edges (arrows), where the real part of the dielectric function crosses zero and the absorption reaches a maximum; and the hybridized phonon-spin-flip excitations. The peaks at the plasma edges both shift and broaden with magnetic field due to a field-induced decrease of the electron mobility, which gives rise to a relative increase of absorption at low energies (vertical red bands) and a decrease around the zero-field plasma edges (blue bands). The hybridized excitations are weaker, but the $OVD_b$ and 0.1% Nb samples display especially clear signatures below the anti-crossing points (diagonal stripes in Fig. 3a,b). Moreover, the results are in quantitative agreement with theoretical calculations of the magneto-absorption based on the model of [4] (Fig. 3d-f), which we adapt to the fact that the soft phonon degeneracy is lifted in the tetragonal phase (see Methods for model and fit details).

The comparison between experiment and theory enables us to estimate the electron-phonon coupling $\tau_R$ (Fig. 3g, including the value for sample $OVD_a$ from above). With the caveat that we ignore the non-sphericity of the Fermi surface and any anisotropy of the Rashba-like coupling, the values at low carrier concentrations are in good agreement with *ab initio* estimates. Yet $\tau_R$ shows a clear increase with electron concentration, which is not expected for Rashba-like coupling within a single band [6]. The lowest-order interband contribution to $\tau_R$, however, should increase in proportion to the square of the Fermi wavenumber [6], which is roughly consistent with Fig. 3g. Our results thus suggest that interband processes become important at concentrations above ~$10^{19}$ $cm^{-3}$. Additional contributions may be present, however, including intraband processes from the upper bands, and a more fine-grained measurement of the concentration dependence should clarify this. The strength of the electron-phonon interaction also strongly depends on the exact atomic displacement pattern of the phonons, which can be expressed as a superposition of a set of distinct phonon eigenvectors [6]. It is therefore possible that the increase of $\tau_R$ is at least partially due to concentration-dependent changes of the phonon eigenvectors. The latter should be discernible, e.g., in detailed inelastic neutron scattering experiments. We further note that phonon anharmonicity is known to be substantial in STO, especially at low carrier concentrations [37,47]; as the softest polar phonon energy increases with carrier density, anharmonic effects decrease, which could influence the electron-phonon coupling as well. The appearance of overtones in the low-concentration $OVD_a$ sample (Fig. 1d) is evidence that nonlinear processes are important at low carrier densities.

Our results have several important implications. First, the finding that the soft polar phonon in insulating STO does not show a measurable magnetic moment in thermodynamic equilibrium complements recent reports of significant magnetic moments in ultrafast measurements [48]. The null result in Fig. 1b implies that the latter must be a non-equilibrium effect, which is relevant to the understanding of chiral phonons in insulators more broadly. Second, the agreement between the measured strength of the Rashba-like EPC and the values obtained from frozen-phonon *ab initio* calculations indicates that the basic physics of the coupling can be captured computationally, at least in the limit in which the phonon frequency lies well below electronic energy scales such as the Fermi energy. The concentration dependence of $\tau_R$ suggests that more detailed *ab initio* studies of interband contributions and anharmonic effects are needed. Moreover, the coupling strengths of about 1000 meV/Å that we obtain above ~$10^{19}$ $cm^{-3}$ are sufficiently large to account for the superconducting transition temperatures for those concentrations. In Fig. 3h, we plot the carrier concentration dependence of the dimensionless interaction parameter $\lambda_{BCS}$ that determines $T_c$ within standard Bardeen-Cooper-Schrieffer (BCS) theory. While $\lambda_{BCS}$ is likely too small to account superconductivity at the lowest carrier densities (which is only observed as filamentary superconductivity in OVD samples [9,49]), the observed values are fully consistent with theoretical estimates [5-7] and conclusions from tunnelling spectroscopy [50] that $\lambda_{BCS}$~0.1-0.2 is needed to explain $T_c$ above $10^{19}$ $cm^{-3}$. The Rashba-like coupling therefore likely plays a role in the superconducting pairing mechanism in STO and related materials with soft polar phonons, including (Pb,Sn,In)Te [19,51] and heterostructures based on $KTaO_3$ [16,18]. More broadly, it might be important for a wide range of other superconductors with polar (but not necessarily soft) phonon branches, especially materials with substantial spin-orbit coupling such as bismuthates [52] and layered van der Waals materials, e.g., $TaS_2$ and $MoS_2$. Of course, it is possible that quadratic electron-phonon interactions are not negligible in STO [39], and that both mechanisms contribute to the pairing, especially at the lowest electron densities where anharmonic effects are more pronounced.

Finally, the hybridization of phonons with electronic spin-flip modes represents a novel route to create chiral phonon-like excitations in metals, and the effective magnetic moment of these excitations is comparable to the electronic moment in the range of fields and energies where the hybridization takes place. In particular, for charge carrier concentrations below ~$10^{19}$ $cm^{-3}$ the $E_u$ mode in STO is so soft that the anticrossing fields are easily reachable with laboratory-scale magnets, and the energies fall into the technologically important millimetre-wave range. The relative field-induced phonon energy shifts in STO are several percent, among the largest known: they are comparable to the phonon shifts in the Dirac semimetal $Cd_3As_2$ induced by a hybridization with electronic cyclotron resonances [24], and nearly an order of magnitude larger than the shifts in magnetic insulators such as $CeF_3$ [53] and materials with strong spin-orbit interactions such as PbTe [25]. Our results also suggest the possibility of chiral phonon engineering, e.g., in STO- and $KTaO_3$-based heterostructures with a magnetic component, plastically deformed STO [54], or closely related materials such as metallic $EuTiO_3$ [55], where the magnetism of $Eu^{2+}$ ions can create significant internal fields. Tunable chiral phonons in materials with high electronic mobility, such as doped STO and related compounds, might thus provide new functionalities in spintronics applications.

**Methods**

*Samples.* High-quality single crystals of STO are obtained commercially (MTI Corp.), and oxygen vacancy doping is achieved through vacuum annealing of insulating samples at 960°C ($OVD_a$) and 1000°C ($OVD_b$) for 2 hours. The carrier density is determined from the Hall coefficient, measured at ambient temperature using a 2 T resistive magnet, except for the $OVD_a$ sample with $2 \cdot 10^{18}$ $cm^{-3}$, where we determine the carrier concentration from the frequency of Shubnikov-de Haas oscillations measured at 1.4 K. Two samples ($OVD_a$, Fig. 1c,d and 0.2% Nb, Fig. 3c) have a gold grid on the surface to provide light polarization along high-symmetry cubic directions, with 3 μm wide and 100 nm thick gold strips deposited using standard lithography techniques at the University of Minnesota Nano Center.

*Far-infrared spectroscopy.* The optical spectroscopy measurements are performed with two different setups. The full spectra shown in Figs. 1 and 3 are obtained with the instrument described in [45], which uses a photomixer diode to generate continuous-wave sub-THz radiation from two near-infrared lasers with controllable wavelengths, and a thermal detection method where the sample is suspended on a sensitive resistive sensor that measures sample temperature changes induced by light absorption. In-house optical spectroscopy measurements (Fig. 1d,e and Fig. 3a) are performed in a vertical 12 T superconducting magnet, using a $^3$He pot with a base temperature of 0.5 K incorporated into the optical probe. For the polarized measurements shown in Fig. 3a, we use a Cu wire grid polarizer placed directly atop the sample. Measurements up to 30 T are performed in a resistive high-field magnet at LNCMI Grenoble, in a pumped liquid helium bath with a base temperature of $T$ = 1.3 K. In order to obtain high signal-to-noise ratios in the noisy high-field environment, it is essential to keep the probe galvanically isolated from the cryostat.

For the polarization-dependent experiments shown in Fig. 2a-c, we employ a 10 T split-coil magnet with optical access (Oxford Instruments), and amplifier-multiplier chains (Virginia Diodes, Inc.) as fixed-frequency sources. Linear polarization is achieved with a wire polarizer, and we use a crosslinked polystyrene Fresnel rhomb for wavelength-independent circular polarization [56]. We demonstrate the efficiency of the rhomb in electron spin resonance measurements on the standard free-radical reference compound 2,2-diphenyl-1-picrylhydrazyl (DPPH) (Supporting Data Figure 1). Optical absorption is measured with the same thermal detection approach as in the broadband experiments.

*Phonon propagator.* The basic Hamiltonian of the dynamic Rashba electron-phonon interaction in a cubic system can be written as [2,4]

$$H_{int} = \Lambda \sum_{\boldsymbol{k},\boldsymbol{q}} \sum_{s,s'} c^{\dagger}_{\boldsymbol{k}+\frac{1}{2}\boldsymbol{q},s} [(\boldsymbol{k} \times \boldsymbol{\sigma}) \cdot \boldsymbol{P}] c_{\boldsymbol{k}-\frac{1}{2}\boldsymbol{q},s'} \quad (1)$$

where $\Lambda$ is a parameter that quantifies the strength of the interaction, $\boldsymbol{k}$ and $\boldsymbol{q}$ are the electron and phonon wavevectors, respectively, $c\dagger$ and $c$ are the electron creation and annnihilation operators, $\boldsymbol{\sigma}$ are Pauli matrices, and $\boldsymbol{P}$ is the phonon polarization, which is proportional to the atomic displacements. In a tetragonal system, the coupling can no longer be quantified with a single number, but with a minimum of three components, and the Hamiltonian must be modified accordingly [6]. We modify the theory of ref. [4] to include coupling to two distinct phonon modes, $A_{2u}$ and $E_u$, but keep all other assumptions; most importantly, we ignore the non-sphericity of the Fermi surfaces, the multiband nature of the STO electronic dispersion, and the possibility that electrons in each band may feature different electron-phonon coupling constants [6]. This is done to reduce the number of free parameters in the fits in Figs. 2c and 3, and the extracted coupling strengths are therefore effective values. Yet for both OVD and Nb-doped samples, the lowest electronic band likely dominates, given its larger density of states []. We also assume a simple parabolic form for the electronic dispersion – while this is an oversimplification, we expect that it does not substantially affect the results, especially those for the OVD samples for which the dispersion anisotropy is not as pronounced [].

A straightforward generalization of the phonon propagator (Eqs. (24-26) in [4]) with magnetic field perpendicular to the phonon polarizations yields for the tetragonal case

$$\begin{aligned} D^{-1} &= D_0^{-1} + \widetilde{\Pi} \\ &= \begin{pmatrix} \omega_{LO}^2 - \omega^2 & 0 & 0 \\ 0 & \omega_{Eu}^2 - \omega^2 & 0 \\ 0 & 0 & \omega_{A2u}^2 - \omega^2 \end{pmatrix} \\ &+ \frac{2}{3} g_0 \begin{pmatrix} -\frac{\Delta^2(\tau_{R,1}^2 - \tau_{R,2}^2)}{\Delta^2 - \omega^2} & 0 & 0 \\ 0 & -\frac{\Delta^2 \tau_{R,1}^2}{\Delta^2 - \omega^2} & \frac{i\Delta\omega\tau_{R,1}\tau_{R,3}}{\Delta^2 - \omega^2} \\ 0 & -\frac{i\Delta\omega\tau_{R,1}\tau_{R,3}}{\Delta^2 - \omega^2} & -\frac{\Delta^2 \tau_{R,3}^2}{\Delta^2 - \omega^2} \end{pmatrix} \end{aligned} \tag{2}$$

where $g_0 = \frac{\sqrt{2} m_b n a_0^2}{\pi^2 \tilde{\rho}}$, with $m_b$ the band mass, $n$ the carrier concentration, $a_0 = 3.9$ Å the lattice parameter, and $\tilde{\rho} = \rho\mu_S/M$ the effective density, where $\rho$ is the mass density and $M$ the molar mass of STO, respectively, and $\mu_S$ is the effective mass of the relevant phonon mode; $\omega_{LO}$, $\omega_{A2u}$ and $\omega_{Eu}$ are the logitudinal optic (LO), $A_{2u}$ and $E_u$ phonon energies, respectively; $\Delta = g\mu_B B$ is the Zeeman energy due to the external magnetic field $B$ (we assume a free-electron gyromagnetic ratio $g = 28.4$ GHz/T); and $\tau_{R,i}$ the Rashba-like electron-phonon coupling constants. For the soft TO modes, only $\tau_{R,1}$ and $\tau_{R,3}$ are relevant, and the LO term does not contribute to the sub-THz optical response, given that the LO phonon energy is ~100 meV [50]. The cubic case is recovered taking $\tau_{R,1} = \tau_{R,3}$ and $\omega_{Eu} = \omega_{A2u}$; this also corresponds to the case where the magnetic field is perpendicular to the tetragonal basal plane, and the light polarization is in that plane. We refer to the electron-phonon coupling constant in that geometry as $\tau_{R,2}$, since it is in principle different from both $\tau_{R,1}$ and $\tau_{R,3}$. For the phonon effective mass, we assume that the TO modes are predominantly Slater modes, where the oxygen cage oscillates in antiphase to the titanium atoms, and $1/\mu_S = 1/m_{Ti} + 3/m_O$. The true displacement pattern of the soft TO phonons is not accurately known, with conflicting conclusions drawn from neutron scattering [57] and hyper-Raman scattering studies [58]. Note though that the effective mass only enters the calculation of $\tau_{R,i}$ from the fits, but it cancels out in the BCS interaction parameter (see below).

The presence of the imaginary off-diagonal terms in the self-energy implies that, for nonzero $\Delta$, the eigenmodes are in general elliptically polarized, i.e. there is a circular component with an effective magnetic moment. The cubic case is particularly simple, since there the eigenstates of the propagator are simply left- and right-circularly polarized modes, and the electron-phonon coupling splits the energies of the two modes in dependence on magnetic field (see also below). The effects of a finite electronic relaxation time are included in the self-energy through a modification of the $\Delta$-dependent terms, Eqs. (D7-D9) in [4].

*Dielectric function and optical absorption.* For light polarization in a [101] plane, we obtain the complex dielectric susceptibility tensor from the phonon self-energy $\widetilde{\Pi}_{ii}$ (in units of energy$^2$) using the relation

$$\chi_{nn} = f\frac{1}{\omega_{nn}^2 - \omega^2 - \widetilde{\Pi}_{nn} - i\omega_{nn}\Gamma_{ph}} \quad (3)$$

where $nn = xx, yy$ are the coordinates in the plane of the polarization, $f$ is the oscillator strength, $\omega_{xx,yy} = \omega_{Eu,A2u}$ and $\Gamma_{ph}$ is a phenomenological phonon damping due to anharmonicity, i.e., unrelated to the electron-phonon interaction. In this case, we neglect the small off-diagonal components of the self-energy, since the $E_u$-$A_{2u}$ splitting is much larger than the electron-phonon coupling term. For the polarization in the basal plane, we work with left ($L$) and right ($R$) circularly polarized eigenmodes, with the susceptibility obtained by taking $nn = L, R$ and $\omega_{ii} = \omega_{Eu}$. In this case, the electron-phonon interaction is the only cause of mode splitting and of course cannot be neglected, but it is straightforward to diagonalize the self-energy in the circular base.

The absorption coefficient is calculated from the dielectric susceptibilities, Eqn. (3), using standard Fresnel formulas, with the addition of a Drude dielectric function to describe the response of the metallic electrons in a magnetic field, $\varepsilon_{Drude}^{L,R} = \frac{i}{\varepsilon_0\omega}\frac{\sigma_{DC}}{1 - i\omega\tau \pm i\mu_H B}$, where $\sigma_{DC}$ is the zero-frequency conductivity, $\tau$ is the electronic relaxation time, and $\mu_H$ is the Hall mobility. For a simple parabolic dispersion with effective mass $m^*$, both the relaxation time and conductivity can be expressed through the Hall mobility as $\tau = m^*\mu_H/e$ and $\sigma_{DC} = ne\mu_H$, respectively, where $e$ is the electron charge and $n$ the carrier density. Based on previous transport results [59], we assume a simple quadratic dependence of the mobility on magnetic field, $\frac{\mu_H(0)}{\mu_H(B)} = 1 + \beta^2 B^2$. The Drude contribution is therefore described by only two parameters, $\mu_H(0)$ and $\beta$, which are both constrained by known transport properties (see below). In order to improve the fits for the OVD and 0.1% Nb samples, we also introduce an effective mass anisotropy of 0.5 and 0.7, respectively, for the Drude contribution along the $x$ and $y$ directions, in line with the realistic dispersion of STO. Importantly, $\tau$ is also used as the inverse electronic damping in the phonon self-energy. In the studied carrier concentration range, the electronic dispersion shows two bands, and a more accurate Drude term would also contain a contribution from the upper band (with a significantly smaller density of states). While we believe that this would further improve the fits, especially for the OVD sample in Fig. 3a, we did not attempt to include the additional term due to the large number of additional free parameters that cannot easily be determined from experiments.

The dominant features in the magneto-optical absorption are the plasma edges, where the absorption coefficient shows a peak, and which strongly broaden in magnetic fields. We also note that cyclotron resonance is included in the Drude form, but due to the relatively large effective masses and field-dependent damping it is not seen in our data, except perhaps in the 0.1% Nb sample at the highest fields (Fig. 3b). This is also not surprising given that the external magnetic field was perpendicular to the sample surface for most measurements, which is a suboptimal arrangement for cyclotron resonance observation.

Circular dichroism (CD) is conventionally defined as

$$\alpha_{CD} = \frac{\alpha_R - \alpha_L}{\alpha_R + \alpha_L} \tag{4}$$

where $\alpha_{R,L}$ are the absorption coefficients for right- and left-circularly polarized light. As noted above, the dominant contribution to the CD response is from domains where the light polarization is in the tetragonal basal plane. We therefore neglect the contribution from the two other types of domains, and calculate the absorption coefficients for each polarization from the susceptibilities in the circular base. The fit in Fig. 2d also contains a multiplicative prefactor as a free parameter, that can be interpreted as the fraction of tetragonal domains with the *c*-axis along the magnetic field. The prefactor also accommodates the fact that the Fresnel rhomb does not yield perfect polarization.

For a given coupling strength $\tau_R$, the BCS parameter in Fig. 3g is obtained from the following relation [5,6],

$$\lambda_{BCS} = \frac{2}{3} N_F \frac{\hbar}{2\mu_S} \frac{(k_F a_0)^2 \tau_R^2}{\omega_{Eu}^2} \tag{5}$$

where $N_F$ is the electronic density of states at the Fermi level, which phenomenologically depends on the carrier concentration as [7,60] $N_F = 0.52 n^{4/3}$ eV$^{-1}$, and $k_F = (3\pi^2 n)^{1/3}$ is the Fermi wavenumber.

*Upper limit of phonon magnetic moment for undoped STO.* When a doubly degenerate phonon shows an intrinsic magnetic moment, the magnetic field lifts the degeneracy and splits the mode into two branches, with energies separated by $\delta = 2\mu_{ph} B$, where $\mu_{ph}$ is the effective phonon magnetic moment [25]. In order to determine the upper limit of $\mu_{ph}$ from our magneto-optical data, we calculate the absorption coefficient for an $E_u$ phonon that splits into two modes at $\omega_\pm = \omega_{Eu} \pm \delta$, and normalize to the zero-field absorption. This ratio is then compared to the RMS noise of the measured absorption ratio at 30 T, which is slightly below 0.01 between 0.3 and 1 meV. If we assume that we can reliably resolve changes that are larger than twice the RMS noise, the smallest $\mu_{ph}$ that would lead to a measurable deviation of the ratio from 1 is approximately $0.01\mu_B$, which represents the upper limit for the moment.

*$E_u$ mode energies and comparison to prior work.* At the lowest studied carrier concentrations, we obtain $E_u$ mode energies that are somewhat lower than what has been reported previously for undoped STO. With the reasonable assumption that the $E_u$ energies monotonically increase with carrier concentration, this implies that in insulating STO this mode is even softer than what was known, which merits comments. We emphasize that, to our knowledge, the $E_u$ energy in metallic STO has not been reported below carrier concentrations of ~$10^{19}$ $cm^3$ at liquid helium temperatures, and the only direct measurement in insulating STO comes from hyper-Raman scattering [46], where it has been estimated as 0.9 meV at ~10 K (with the possibility that the actual sample temperature was even higher due to laser-induced sample heating). Inelastic neutron scattering experiments typically do not resolve the $E_u$ and $A_{2u}$ modes, and therefore obtain significantly higher effective energies for the TO phonon (1.6-1.8 meV at low temperatures) [37,61,62]. A recent cold neutron scattering study [63] does show evidence of both modes, but the $E_u$ phonon is not resolved clearly due to the influence of Bragg peak tails at energy transfers below 1 meV. Given that the $E_u$ energies determined from hyper-Raman scattering do not saturate in the temperature range studied with that technique, it is plausible that the phonon softens further in cooling to sub-Kelvin temperatures.

*Fitting and constraints.* The electron-phonon coupling strength for the OVD sample in Fig. 1d,e is obtained from a simple fit of the eigen-energies of the cubic propagator to the magnetic field dependence of the minima in the magneto-optical spectra. As free parameters, we use the zero-field $E_u$ mode energy and the coupling $\tau_2$. The effective mass is taken to be $2.5m_e$ based on prior specific heat and quantum oscillation measurements [60,64]. We check the consistency of the fit by extracting the value of $\tau_2$ independently from the frequency difference between the two phonon branches close to the anticrossing field (5 T or 0.6 meV), seen as a minimum and a maximum in the raw data in Fig. 1e. We use Eq. (65) in [4] (with the correct effective density for a Slater mode), and obtain a value of $\tau_2$ similar to the fit.

For the second OVD sample (Fig. 2 and Fig. 3a), we first perform a fit to the CD data (Fig. 2d) to obtain the phonon width $\Gamma_{ph}$, the coupling $\tau_2$, and the zero-field Hall mobility. In this fit, we constrain the parameter $\beta$ to be below 1.3 $meV^{-1}$, based on published magnetoresistance data [59], and the $E_u$ phonon energy is constrained to be between 0.85 and 0.95 meV, based on the position of anticrossing features in Fig. 3a. The parameters from the CD fit are then used to model the magneto-optical data in Fig. 3a, with the relative fractions of the three types of domains, the $A_{2u}$ phonon energy, and the oscillator strength $f$ as free parameters. Note that the ratio of intensities of the plasma edge and phonon features cannot be brought into complete agreement with the data, likely due to the oversimplification of the Drude contribution, i.e., the omission of the second band, as noted above. Yet it is also possible that the intensity change is due to a decrease of the phonon width above the anticrossing point, an effect not captured in the lowest-order theory and an interesting topic for further work. For this sample, we also perform a check of the value of $\tau_2$ by calculating the coupling from the mode splitting close to the anticrossing field (7.7 T or 0.9 meV in this case).

In the magneto-optical data for the 0.1% Nb sample (Fig. 3b), two screened plasma edges (due to the different energies of the $E_u$ and $A_{2u}$ modes which provide the screening) overlap substantially with the phonons, and represent the dominant features in the data. Nevertheless, the hybridized phonon-spin-flip excitation is clearly visible, and it is possible to perform fits to the optical absorption obtained from the full tetragonal dielectric function. The free parameters in this case are the Hall mobility, the oscillator strength $f$, the $E_u$ and $A_{2u}$ mode energies, the phonon width, the magnetoresistance parameter $\beta$, and the electron-phonon coupling strengths. To reduce the number of free parameters, we assume that all $\tau_i$ are equal. The results for this sample are obtained using unpolarized light, and we assume an equal population of all three types of domains. The angular averaging likely also affects the value of the extracted effective electron-phonon coupling, which is predicted to be anisotropic [6,8]; we expect that the average of $\tau_i$ over the Fermi surface is smaller than the maximum values, and the extracted coupling is therefore an underestimate. All parameters obtained from the fit are in good agreement with prior transport and optical spectroscopy results.

Data for the sample with 0.2% Nb shown in Fig. 3c are analysed taking into account the fact that this sample has a micrometer gold grid deposited on the surface. We use a simple model for the dielectric function of the gold layer, similar to standard models of wire polarizers [65]: for light polarized along the gold strips, the dielectric function is obtained as a parallel connection between the gold and vacuum:

$$\varepsilon_{gold}^{TE} = \frac{1}{2}\left(1 + \frac{i\sigma_{gold}}{\varepsilon_0 \omega}\right) \tag{6}$$

with $\sigma_{gold}$ the conductivity of the gold, which is expected to be significantly larger than in the bulk due to the small thickness of the layer. For light polarized perpendicular to the strips, we use a series connection instead,

$$\varepsilon_{gold}^{TM} = \frac{2}{1 + \frac{\varepsilon_0 \omega}{i\sigma_{gold}\varepsilon_0 \omega}}. \tag{7}$$

The total reflectivity of the sample with the gold layer on the surface is then calculated from standard Airy formulas for thin films, and the absorptivities for both polarization directions are added. The fit in Fig. 3f is obtained assuming that the $E_u$-$A_{2u}$ degeneracy is negligible for this carrier concentration, i.e., we use the cubic self-energy, since the data do not show sufficiently clear phonon and/or plasma edge features to separate the two modes. Similar to the 0.1% Nb sample analysis, the free parameters in the fit are the Hall mobility, oscillator strength, phonon width, $\beta$ and $\tau_2$, with an additional free parameter, the gold conductivity $\sigma_{gold}$. We note that the Hall mobility is constrained to be between 2000 and 4000 $cm^2V^{-1}s^{-1}$, based on prior transport results on similar samples [28,29,35]. Although the fit is not very sensitive to $\sigma_{gold}$, the best-fit value is $\sigma_{gold} \sim 3 \cdot 10^5$ S/cm, which is reasonable for a 100 nm thick and micron-wide strip [66].

**Acknowledgements**

We thank A. Garcia-Etxarri, P. Volkov, A. Kumar, P. Chandra, J. Ruhman, S. Kumar Saha, V. Kataev, M. Požek, S. Hameed, M. Minola, and A. McLeod for helpful comments and discussions.

**Funding statement**

The work in Zagreb was supported by the Croatian Science Foundation through Grant No. UIP-2020-02-9494, using equipment funded in part through project CeNIKS co-financed by the Croatian Government and the European Union through the European Regional Development Fund–Competitiveness and Cohesion Operational Programme (Grant No. KK.01.1.1.02.0013). NS acknowledges support from the Croatian Science Foundation mobility program, Grant No. MOBDOK-2023-4419. The work at the University of Minnesota was funded by the US Department of Energy through the University of Minnesota Center for Quantum Materials, under Grant No. DE-SC-0016371. Portions of this work were conducted in the Minnesota Nano Center, which is supported by the National Science Foundation through the National Nanotechnology Coordinated Infrastructure (NNCI) under Award Number ECCS-2025124. The work in Dresden was supported by the Deutsche Forschungsgemeinschaft (DFG, German Research Foundation) through Grant no. 499461434 (AL 1771/8-1) and the Dresden-Würzburg Cluster of Excellence (EXC 2147) "Complexity, Topology and Dynamics in Quantum Matter (ctd.qmat)" (project-id 390858490). MNG is supported by the Ramon y Cajal Fellowship RYC2021-031639-I funded by MCIN/AEI/ 10.13039/501100011033. AvK acknowledges support by the Israel Science Foundation (ISF) grant No. 3152/25, and by the United States – Israel Binational Science Foundation (BSF), grant No. 2022242.

**Author contributions**

DP, AvK and MNG designed the study. IK, NS and MG prepared samples; NS, LR, IK, JD, FLM, MO, BB, AA and DP performed optical spectroscopy measurements and analysed data; NS, IK, DP, ET and MB performed transport measurements; AnK, AvK and MNG performed theoretical calculations. DP wrote the paper, with input from all authors.

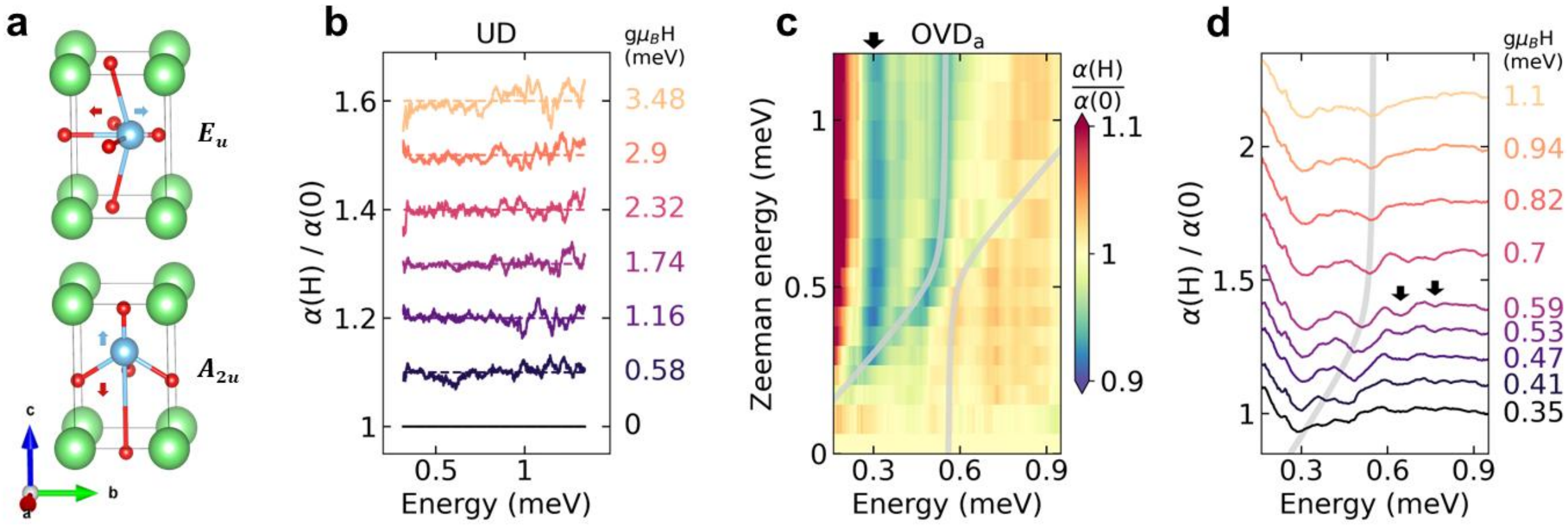

**Figure 1 | Magnetic field effects on soft polar phonons in STO**. **a**, Sketch of atomic displacements for the two softest transverse optic polar phonon modes in the tetragonal phase of STO, with $E_u$ and $A_{2u}$ symmetry; the displacements correspond to antiphase motion of the titanium atom and oxygen cage (Slater mode). **b**, Magneto-optical absorption across the $E_u$ phonon in undoped (UD) insulating STO in magnetic fields up to $H = 30$ T, at 1.3 K and normalized by the zero-field absorption; the field is expressed as the Zeeman energy $E_Z = g\mu_B H$ using the free-electron gyromagnetic ratio $g = 2$. We do not observe any changes within the noise, which puts an upper limit on the phonon magnetic moment of $\sim 0.01\mu_B$. (**c**) Magneto-optical absorption in a metallic STO sample (OVD$_a$) with charge carrier concentration $2 \cdot 10^{18}$ cm$^{-3}$, at 0.7 K. A hybridized phonon-spin-flip excitation is seen around photon energy 0.5 meV, demonstrating the presence of a spin-orbit-mediated electron-phonon coupling in excellent agreement with theoretical predictions (grey lines). The additional vertical line around 0.3 meV is a plasma absorption edge (black arrow). (**d**) Raw magneto-optical spectra that more clearly show additional signals above the main phonon branch (arrows). The features are consistent with overtones, most likely due to interactions between phonons, electronic spin flip modes and plasma oscillations. The data are shifted vertically for clarity. Grey line: low-energy branch of the hybridized spin-phonon excitation.

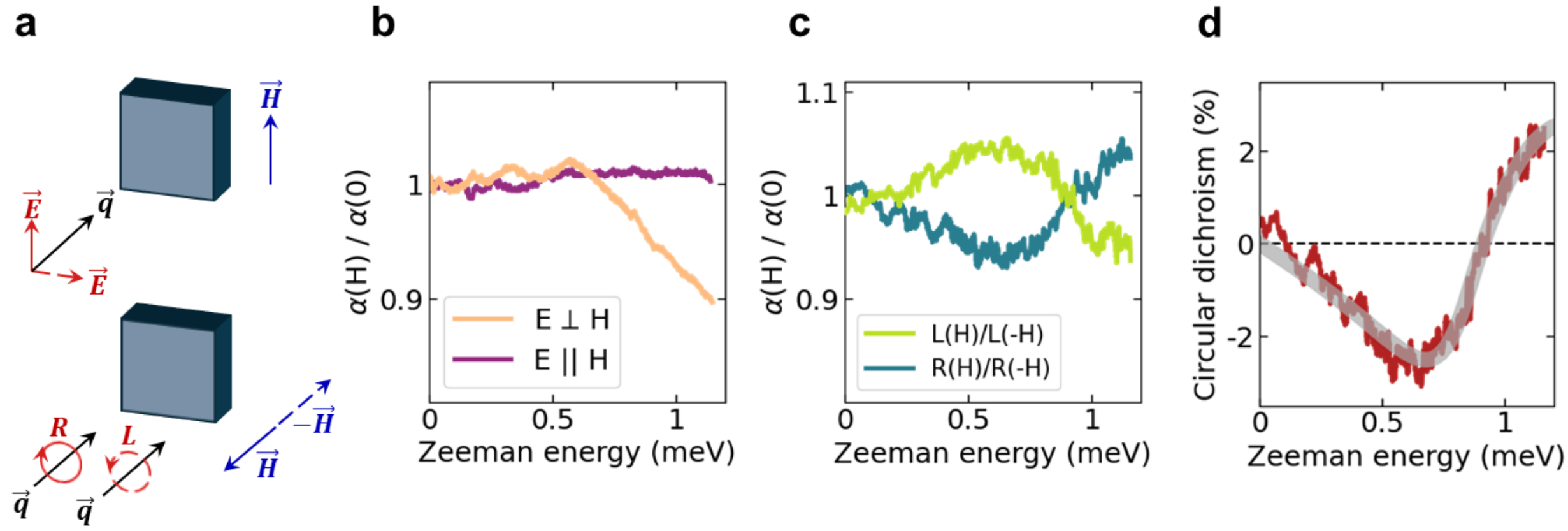

**Figure 2 | Effects of light polarization on the magneto-optical response of metallic STO. a**, Sketch of the experimental geometry for linear and circularly polarized light. **H** is the external magnetic field, **q** is the light propagation direction, **E** the electric field, and *R* and *L* denote left- and right-hand circularly polarized light. **b**, Magneto-optical absorption for light linearly polarized perpendicular and along the direction of the magnetic field, measured at photon energy 1 meV for sample $OVD_b$ with carrier concentration $7.2 \cdot 10^{18}$ $cm^{-3}$. A field-induced change is only observed in the perpendicular direction, in line with theoretical predictions. **c**, Absorption of circularly polarized light in dependence on magnetic field at 1 meV. A change of handedness inverts the signal, as expected in a system with field-induced chiral response. **d,** Magnetic circular dichroism (CD) calculated from the data in (**c**) assuming 100% polarization efficiency. The grey line is a fit to the theoretical dependence (see Methods for details).

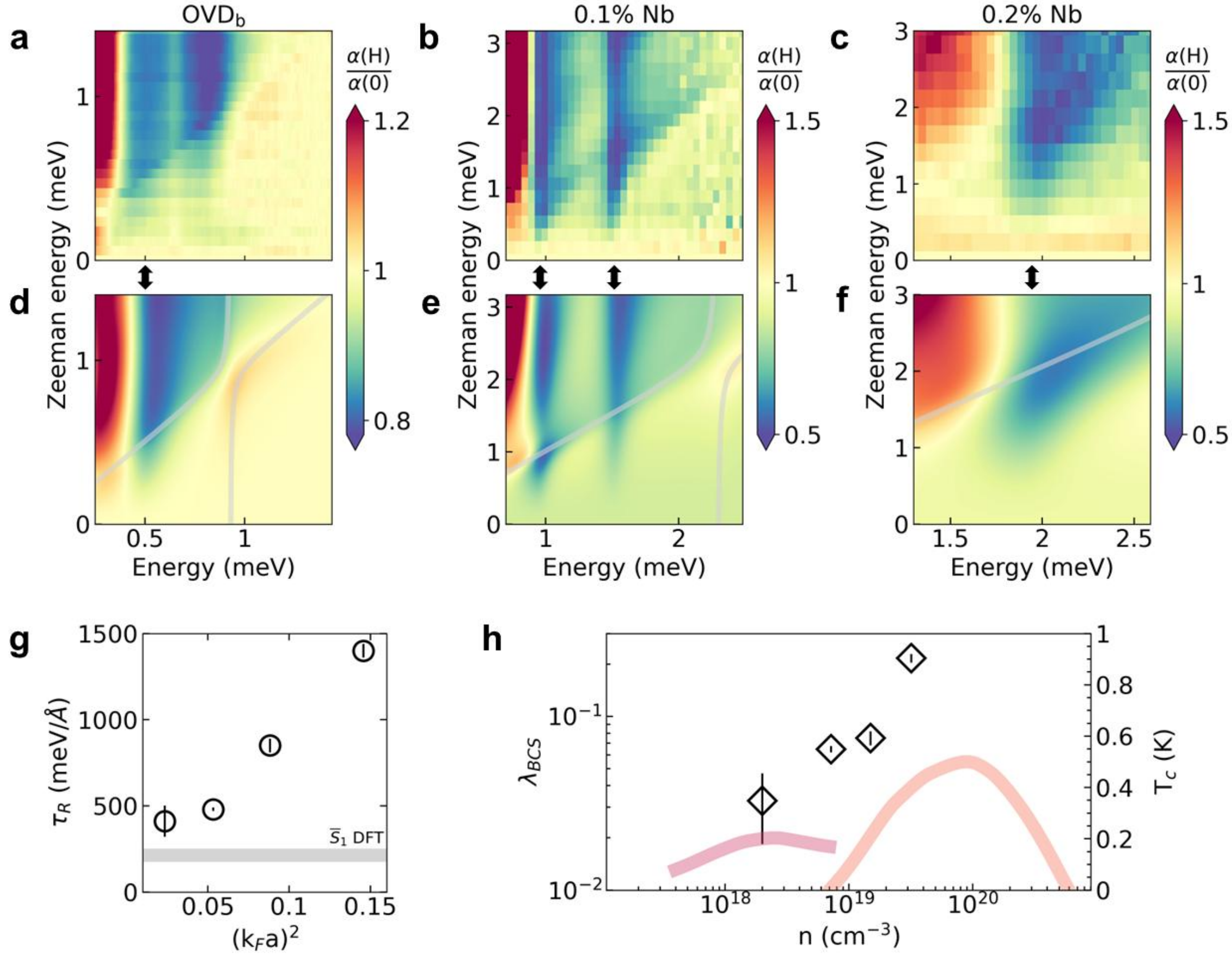

**Figure 3 | Doping dependence of the magneto-optical response of metallic STO. a-c**, Data for samples with three different electron concentrations: $7.2 \cdot 10^{18}$ cm$^{-3}$ (OVD$_b$), $1.5 \cdot 10^{19}$ cm$^{-3}$ (0.1% Nb-Ti substitution) and $3.2 \cdot 10^{19}$ cm$^{-3}$ (0.2% Nb-Ti substitution). In all cases, the hybridized phonon-spin-flip chiral excitations partially overlap with plasma absorption edges (blue vertical features around 0.5 meV in (a), 1 and 1.6 meV in (b) and 2 meV in (c)), but they are clearly visible as sloped lines below the anticrossing point, especially in (a) and (b). **d-f**, Colormaps: fits of theoretical magneto-optical absorption to the data in (a-c), showing excellent agreement (see Methods for details of the fit procedure). Vertical arrows: energies of plasma edges; grey lines: hybridized spin-phonon excitation energies obtained from the fit parameters. **g**, The Rashba-like electron-phonon coupling $\tau_R$ extracted from the fits in (d-f) and Fig. 1c, vs. the square of the Fermi wavenumber. The approximately linear dependence indicates that interband processes become important at higher electron densities. Horizontal line: *ab initio* density functional theory (DFT) calculation for intraband coupling to a Slater mode [6]. **h,** Dimensionless interaction parameter $\lambda_{BCS}$ which determines $T_c$, in dependence on the electron concentration, along with the experimental superconducting phase diagram (lines). Above $n \sim 10^{19}$ cm$^{-3}$, $\lambda_{BCS}$ is sufficiently large to explain $T_c$.

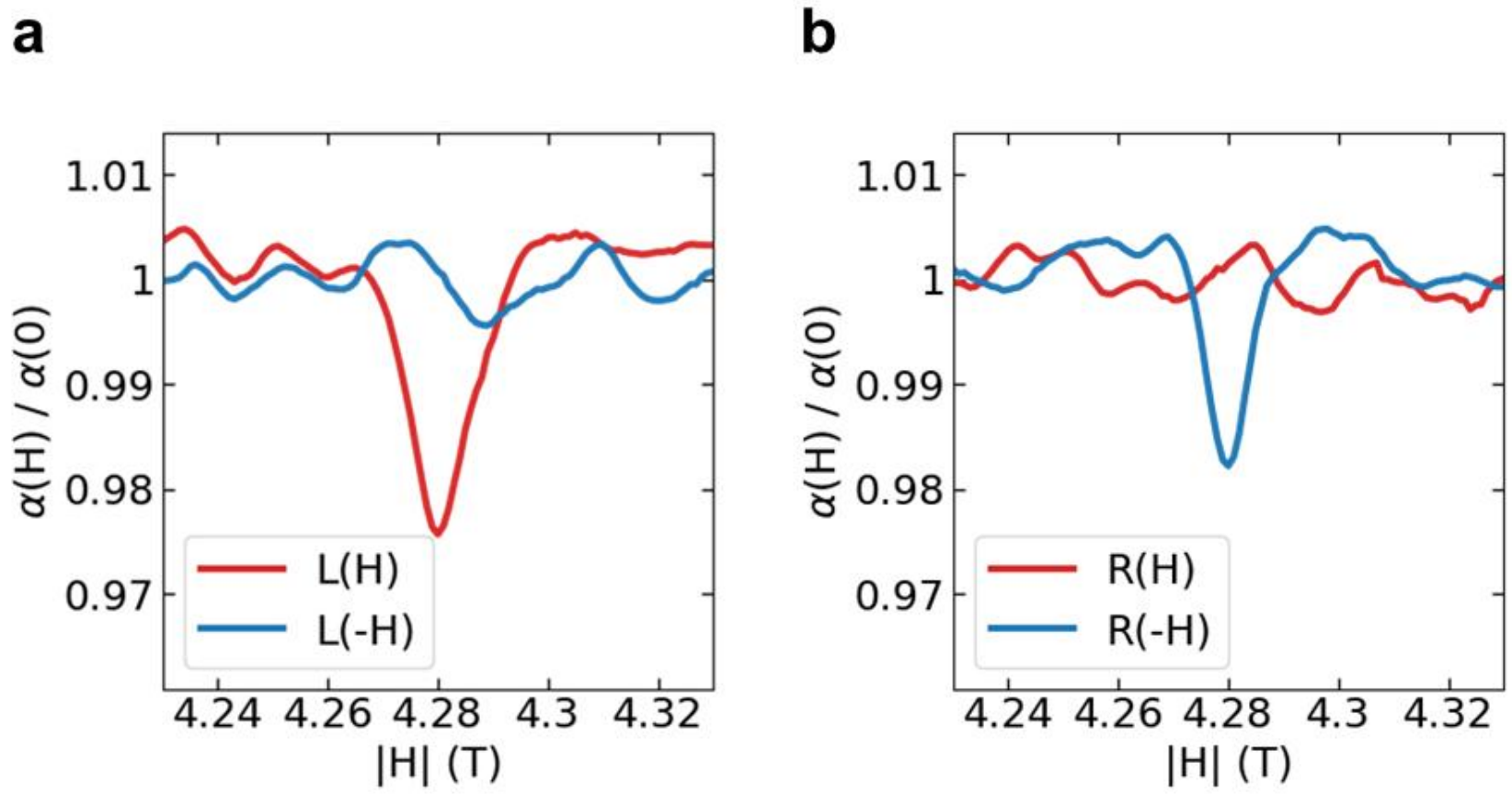

**Extended Data Fig. 1 | Demonstration of circular dichroism in the reference free-radical compound 2,2-diphenyl-1-picrylhydrazyl (DPPH). a** and **b** show the magneto-optical absorption for left- and right-circularly polarized light, respectively, in a small (80 μg) crystal of DPPH at 2 K. The electron spin resonance absorption line only appears in positive fields for L-polarization and in negative fields for R-polarization, which demonstrates the high efficiency of the Fresnel rhomb polarizer.